\documentclass[journal,twocolumn]{IEEEtran}

\newcommand{\pf}{proof} 
\usepackage{amsthm}

\usepackage{amssymb, amsmath, xcolor, graphicx, enumerate, cite}

\usepackage{tikz}
\usepackage{caption}
\usepackage{subcaption}

\usepackage{algorithm}
\usepackage[noend]{algpseudocode}

\newcommand{\bi}{\begin{itemize}}
\newcommand{\ei}{\end{itemize}}
\newcommand{\vo}[1]{\boldsymbol{#1}}
\newcommand{\x}{\vo{x}}

\newcommand{\X}{\vo{X}}

\newcommand{\y}{\vo{y}}

\newcommand{\K}{\vo{K}}

\newcommand{\Q}{\vo{Q}}

\newcommand{\R}{\vo{R}}
\newcommand{\Y}{\vo{Y}}
\renewcommand{\H}{\vo{H}}
\newcommand{\I}{\vo{I}}
\newcommand{\M}{\vo{M}}

\newcommand{\Z}{\vo{Z}}
\newcommand{\z}{\vo{z}}

\newcommand{\Xbar}{\vo{\bar{X}}}

\newcommand{\D}{\vo{D}}
\newcommand{\C}{\vo{C}}
\newcommand{\A}{\vo{A}}
\newcommand{\B}{\vo{B}}
\newcommand{\G}{\vo{G}}
\newcommand{\n}{\vo{n}}

\newcommand{\N}{\vo{N}}
\newcommand{\w}{\vo{w}}
\newcommand{\W}{\vo{W}}

\newcommand{\sA}{\boldsymbol{\mathcal{A}}}
\newcommand{\sB}{\boldsymbol{\mathcal{B}}}
\newcommand{\sC}{\boldsymbol{\mathcal{C}}}
\newcommand{\sD}{\boldsymbol{\mathcal{D}}}
\newcommand{\sQ}{\boldsymbol{\mathcal{Q}}}
\newcommand{\sR}{\boldsymbol{\mathcal{R}}}
\newcommand{\sS}{\boldsymbol{\mathcal{S}}}
\newcommand{\sK}{\boldsymbol{\mathcal{K}}}
\newcommand{\sG}{\boldsymbol{\mathcal{G}}}
\newcommand{\sH}{\boldsymbol{\mathcal{H}}}
\newcommand{\Gamb}{\boldsymbol{\Gamma}}

\renewcommand{\P}{\vo{P}}

\newcommand{\blue}[1]{\textcolor{blue}{#1}}

\newcommand{\real}{\mathbb{R}}
\newcommand{\Exp}[1]{\mathbb{E}\left[#1\right]}
\newcommand{\diag}[1]{\mathbf{diag}\left(#1\right)}

\newcommand{\Var}[1]{\Exp{#1{#1}^T}}
\newcommand{\mup}{\vo{\mu}^{-}}

\newcommand{\Sig}{\vo{\Sigma}}
\newcommand{\mupp}{\vo{\mu}^{+}}

\newcommand{\Sigpp}{\vo{\Sigma}^{+}}

\newcommand{\mub}{\vo{\mu}}
\newcommand{\Sigb}{\vo{\Sigma}}

\newcommand{\trace}[1]{\mathbf{tr}\left(#1\right)}

\newtheorem{theorem}{Theorem}

\theoremstyle{remark}
\newtheorem{remark}{\textbf{Remark}}

\newcommand{\eqnlabel}[1]{\label{eqn:#1}}
\newcommand{\figlabel}[1]{\label{fig:#1}}
\newcommand{\eqn}[1]{(\ref{eqn:#1})}
\newcommand{\fig}[1]{fig.(\ref{fig:#1})}
\newcommand{\Fig}[1]{Fig.(\ref{fig:#1})}

\begin{document}

\title{Optimal Sensor Precision for Multi-Rate Sensing for Bounded Estimation Error}
\author{Niladri Das and Raktim Bhattacharya%
\thanks{Niladri Das (graduate student) and Raktim Bhattacharya (associate professor) are with the Department
of Aerospace Engineering, Texas A\& M University, College Station,
TX, 77845 USA. E-mail: \texttt{niladri@tamu.edu, raktim@tamu.edu.}
}}

\maketitle

\begin{abstract}
We address the problem of determining optimal sensor precisions for estimating the states of linear time-varying discrete-time stochastic dynamical systems, with guaranteed bounds on the estimation errors. This is performed in the Kalman filtering framework, where the sensor precisions are treated as variables. They are determined by solving a constrained convex optimization problem, which guarantees the specified upper bound on the posterior error variance. Optimal sensor precisions are determined by minimizing the $l_1$ norm, which promotes sparseness in the solution and indirectly addresses the sensor selection problem. The theory is applied to realistic flight mechanics and astrodynamics problems to highlight its engineering value. These examples demonstrate the application of the presented theory to a) determine redundant sensing architectures for linear time invariant systems, b) accurately estimate states with low-cost sensors, and c) optimally schedule sensors for linear time-varying systems.
\end{abstract}


\section{Introduction}
Kalman filtering provides the state estimate with minimum error covariance for a given sensor precision. Sensor precision is defined as the inverse of sensor noise. In this paper, we look at the inverse problem. 
Given a dictionary of sensors, what is the most noise in these sensors such that the error covariance is below a given error bound. This problem is of significant engineering value since it allows system designers to determine the optimal accuracy of sensing components that satisfy a given system-level error budget. Since cost of sensors are proportional to the accuracy, the problem has economical implications as well. In this paper we present convex optimization formulations to solve the optimal sensing precision problem in a Kalman filtering framework, for linear discrete-time time-varying systems with multi-rate sensing.

The proposed framework also impacts other important problems in sensing, such as sensor scheduling and selection. Existing algorithms for sensor scheduling \cite{Guestrin_2005, Gupta_2006, Zhang_2006, He_2006, Shi_2013, Pequito_2013, Jawaid_2015, Summers_2016, Han_2017, Chen_2017} and selection \cite{Joshi_2009, Dhingra_2014, Chepuri_2015, Tzoumas_2016, zhang2017sensor} assume that the sensor’s noise variance is given. The framework in this paper can be used to determine them optimally.

While this paper focuses on determining the optimal sensor precisions, the presented framework can also be applied to schedule and select sensors. Starting with a dictionary of sensors with unknown precisions, minimizing the $l_1$ norm of the sensor precisions will promote sparseness in the solution. Sensors with zero precisions can be eliminated from the system, and sensors with non zero precisions will guarantee the required estimation accuracy. Therefore, it is possible to simultaneously determine the optimal sensor precisions and prune out unnecessary sensors using the proposed framework. This is appealing because the problem can be solved in a convex optimization framework in a general setting. This is in contrast with the NP-hard formulations and heuristic methods to solve them. 

\subsection{Key Contribution}
The problem of determining optimal sensor precision was first introduced in \cite{LI_2008}, in the context of output-feedback controller design for continuous-time systems without uncertainty and steady-state performance guarantees. Recently, we extended that work to determine optimal sensing precisions for continuous-time robust output-feedback control, with guaranteed $\mathcal{H}_2$ performance \cite{saraf2017}. 

In this paper, we look at the problem of determining the optimal sensor precision for state estimation of linear time-varying, discrete-time, stochastic systems with multi-rate sensing. To the best of our knowledge, this is the first paper that determines optimal sensor precision for these systems. The main results are presented as two theorems, deriving the convex optimization formulations, which achieve the objective mentioned above. The first theorem addresses the problem of determining optimal sensor precisions to bound the estimation error over one time step. The second theorem determines the optimal sensor precision to bound the steady-state error for periodic time-varying systems. 

\subsection{Layout}
Section \ref{sec:preliminaries} presents the preliminaries for this paper, where we define the model of the dynamical system and the sensor model assumed in the paper. We also define the problem to determine the optimal precisions for a given upper bound in the estimation error. Section \ref{sec:oneTime}  derives the optimization problem for determining the sensing precision if the objective is to bound the error after a one-time step.  Section \ref{sec:steady-state}, derives the optimization problem for determining the sensor precisions to bound the steady-state estimation error for periodic systems. Section \ref{sec:examples} highlights the engineering relevance of the proposed framework, where the theoretical results are applied to state-estimation problems from flight mechanics and astrodynamics. In that section, we present three examples that demonstrate how the proposed theory can be applied to identify redundancy in sensing,  accurately estimate states with cheap sensors, and optimally schedule sensors for time-varying periodic systems.

\section{Preliminaries}
\label{sec:preliminaries}
We focus on determining the optimal sensor precision for linear time-varying discrete-time stochastic systems described by the model of the form:
\begin{subequations}
\begin{align}
\x_{k+1} &= \A_{k}\x_{k} + \B_{k}\w_{k}, \eqnlabel{processDynamics}\\
\y_k &= \C_k\x_k + \n_k, \eqnlabel{sensing}
\end{align}\eqnlabel{stochasticSystem}
\end{subequations}
where $k=0,1,2,...$ are the time indices,  $\A_{k}\in \real^{n_x\times n_x}, \B_{k}\in\real^{n_x\times n_w}$ and $\C_k\in\real^{n_{y_k}\times \n_x}$ are the system matrices, $\x_k\in\real^{n_{x}}$ is the $n_x$ dimensional state of the \textit{model} at time instant $k$, $\w_k \in\real^{n_w}$ is the $n_w$ dimensional zero-mean Gaussian additive process noise variable with \sloppy $\mathbb{E}[\w_k\w_l^T] = \Q_k$, where $\mathbb{E}[.]$ denotes the expected value. The $n_{y_k}$ dimensional observations at time $k$ is denoted by $\y_k\in\real^{n_{y_k}}$, which is corrupted by an $n_{y_k}$ dimensional additive observation noise $\n_k \in\real^{n_{y_k}}$ at time instant $k$. The sensor noise at each time instant is a zero mean Gaussian random variable with $\mathbb{E}[\n_k\n_l^T] = \R_k\delta_{kl}$. The initial conditions are $\Exp{\x_0}=\boldsymbol{\mu}_{0}$ and $\Var{(\x_0-\mub_0)} = \Sig_0$. The process noise $\w_{k}$, observation noise $\n_k$, and initial state variable $\x_0$ are assumed to be independent. 

In Kalman filtering, the propagation equations for the mean and the error covariance are given by \cite{anderson2012optimal}
\begin{align*}
\mup_{k} = \A_k\mupp_{k-1} & \text{ and } \Sig_k^- = \A_k\Sig^{+}_{k-1}\A_k^T+\B_k\Q_k\B_k^T,
\end{align*}
and update equation for mean is $\mupp_{k} =\mup_{k} + \K_k(\y_k-\C_k\mup_{k})$, and for variance is
\begin{equation}
\Sig^{+}_{k}= (\I_{n_x}-\K_k\C_k)\Sig_k^-(\I_{n_x}-\K_k\C_k)^T + \K_k\R_k\K_k^T,
\end{equation}
where $\K_k$ is the Kalman gain determined by minimizing $\trace{\Sig^{+}_{k}}$ for a given $\R_k$. 

In this paper, we treat $\R_k$ as a \textit{variable}, and for a given error bound $\gamma_d\in \real^+$ (positive real number) we determine the \textit{maximum} $\R_k$ such that $\trace{\Sigpp_k} \le \gamma_d$. Assuming $\R_k$ to be diagonal, i.e. $\R_k := \diag{\vo{r}_k}$, where $\vo{r}_k := \begin{bmatrix} r_1 & r_2 & \cdots & r_{n_{y_k}} \end{bmatrix}^T$, with $r_i > 0$.  As shown later, it is convenient to formulate the problem in terms of sensor precisions, defined by $\vo{S}_k$, which is the inverse of sensor noise $\R_k$, i.e. $\vo{S}_k:=\R_k^{-1}$, resulting in $s_i:=1/r_i$. Defining $\vo{s} := \begin{bmatrix} s_1 & s_2 & \cdots & s_{n_y} \end{bmatrix}^T$, we can determine the least precise sensors, i.e. minimize $\trace{\vo{S}_k}$, for which 
$\trace{\Sigpp_k} \le \gamma_d$ is guaranteed. Since more precise sensors are more expensive, satisfying required accuracy with least precise sensors has favorable economic implications. 

Minimization of $\trace{\vo{S}_k}$ also has sparseness implications as $\trace{\vo{S}_k}$, for $\vo{S}_k \ge 0$, is equivalent to the $\|\vo{s}\|_1$. Since it is well-known that the $l_1$ norm is sparseness promoting \cite{hastie2009elements}, minimizing $\trace{\vo{S}_k}$ will result in a sparse solution that satisfies $\trace{\Sigpp_k} \le \gamma_d$, if a sparse solution exists for the problem. Consequently, sensors with zero precisions would not contribute to achieving $\trace{\Sigpp_k} \le \gamma_d$, and thus can be removed from the system. It should also me mentioned that for a given system parameters, there exists a range of $\gamma_d$, for which the minimization problem is infeasible.

With this background, we next present the formulation to determine the optimal precision for multi-rate information fusion in the Kalman filtering framework.
 
\section{Optimal Sensing Precision for Update after One Time Step} \label{sec:oneTime}
Here we present the formulation that determines the optimal sensing precision that guarantees a bounded estimation error after one update. The problem is formulated in a batch processing framework, where $m$ measurements are collected before the state is updated. Specifically, given the state uncertainty at time $t_{km}$, the objective is to determine the precisions of these $m$ measurements, such that the state estimate at time $t_{(k+1)m}$ has bounded error.

Consider sensing over $m$ time steps, as shown below.
\begin{figure}[h!] \centering
\begin{center}
\begin{tikzpicture}
\draw[thick] (0,0) -- (3,0);
\draw[dotted,thick] (3,0) -- (5,0);
\draw[thick] (5,0) -- (6,0);

\draw[fill] (0,0) circle (1mm) node[below=3mm] {$t_{km}$} node[above=3mm]{$\vo{\blue{x}}_{\blue{km}}$};
\draw[fill] (1,0) circle (1mm) node[below=3mm] {$t_{km+1}$} node[above=3mm]{$\vo{y}_{km+1}$};
\draw[fill] (2,0) circle (1mm) node[below=3mm] {$t_{km+2}$} node[above=3mm]{$\vo{y}_{km+2}$};
\draw[fill] (3,0) circle (1mm);
\draw[fill] (5,0) circle (1mm);
\draw[fill] (6,0) circle (1mm) node[below=3mm] {$t_{(k+1)m}$} node[above=3mm]{$\begin{matrix}\vo{\blue{x}}_{\blue{(k+1)m}} \\[2mm] \vo{y}_{(k+1)m} \end{matrix}$};
\end{tikzpicture}
\caption{Multi-rate measurements over $m$ time steps.}
\figlabel{mr:sensor}
\end{center}
\end{figure}
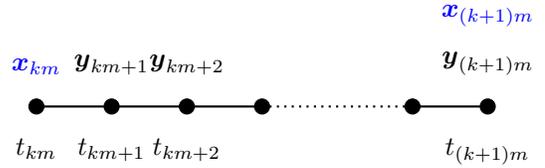
This allows us to model multi-rate sensing, with $m$ being the least-common-multiple of the various sensing intervals. With each measurement $\vo{y}_{km+j}$, for $j=1,\cdots,m$, we associate sensor noises $\vo{r}_{km+j} \in \real^{n_{y_{km+j}}}$. 

In conventional Kalman filtering, the sensor noises $\vo{r}_{km+j}$ are known and the objective is to estimate the state at time $t_{(k+1)m}$, given the posterior at time $t_{km}$ and measurements $\vo{y}_{km+1},\cdots,\vo{y}_{(k+1)m}$, at times $t_{km+1},\cdots,t_{(k+1)m}$. This scenario is common in control system applications where the control-loop is band limited (to prevent excitation of high-frequency dynamics), but the sensing loop can be faster. In such a scenario, the Kalman filter determines the state estimate every $m$ time steps, by batch-processing $m$ measurements. In this paper, we are interested in maximizing $\vo{r}_{km+j}$ for which the posterior state estimation error at time $t_{(k+1)m}$ satisfies a given upper bound.

\subsection{Augmented Dynamical System} 
To determine the posterior at time $t_{(k+1)m}$, we need to propagate the state uncertainty from $t_{km}$ to $t_{(k+1)m}$, to obtain the prior at $t_{(k+1)m}$. This is done by lifting discrete-time signals defined over times $t_k$ to signals defined over times $t_{km}$, for $k=0,1,\cdots,\infty$. We define lifted signals $\X_{k}\in\real^{mn_x}$ and $\Y_{k}\in\real^{n_{y_{k,m}}}$ as a vector with $m$ consecutive state vectors and measurements stacked vertically, respectively, i.e. 
\begin{align}
 \X_{k} := \begin{pmatrix}
 \x_{km+1} \\ \vdots \\ \x_{(k+1)m}
 \end{pmatrix}, & \Y_{k} := \begin{pmatrix}
		\y_{km+1} \\ \vdots \\ \y_{(k+1)m}
	\end{pmatrix},
\end{align}
where $m\geq 1$, and $n_{y_{k,m}} = \sum_{j=1}^m n_{y_{km+j}}$.

Using the system defined in equation \eqn{processDynamics}, the state $\X_k$ can be expressed as \cite{koch2015accumulated}
\begin{equation}
	\X_{k}  = \sA_k\x_{km} + \sB_k\W_{k}, \eqnlabel{augDyn}
\end{equation}
where
\begin{align}
\sA_k &:= \begin{bmatrix}\A_{km} \\ \A_{km+1}\A_{km} \\ \vdots \\ \displaystyle\prod_{i=0}^{m-1}\A_{km+i}\end{bmatrix},\\
\sB_k &:= \begin{bmatrix}\B_{km} & 0 & ... & 0\\
\A_{km+1}\B_{km} & \B_{km+1} & ... & 0\\
\vdots & \vdots & & \vdots\\
\displaystyle\prod_{i=1}^{m-1}\A_{km+i}\B_{km} & ... & ... & \B_{km+m-1}
\end{bmatrix}, \\
\W_{k} &:= \begin{pmatrix} \w_{km} \\ \vdots \\ \w_{km+m-1}\end{pmatrix}, 
\end{align}
with the product terms in $\sA_k$ and $\sB_k$ defined as
\begin{align}
 \prod_{i=i_1}^{i_2}\A_{k+i} := \A_{k+i_2}\times\cdots\times\A_{k+i_1+1}\A_{k+i_1} \text{if $i_2 \geq i_1$}.
\end{align}
The augment measurement model from \eqn{sensing} is
\begin{align}
\Y_{k} &= \sC_k\X_k + \N_{k}, \eqnlabel{augY}
\end{align}
where
\begin{align}
\sC_k &:= \diag{\C_{km+1},...,\C_{(k+1)m}},\eqnlabel{Ck}
\end{align}
 and
\begin{align}
\N_{k} &:= \begin{pmatrix} \n_{km+1}\\ \vdots \\ \n_{(k+1)m}\end{pmatrix}.
\end{align}
We define the augmented process noise and observation noise variances as:
\begin{align}
\nonumber &\sQ_k := \Var{\W_{k}} = \diag{\Q_{km},\cdots,\Q_{km+m-1}},
\end{align}
and
\begin{align}
&\sR_k := \Var{\N_{k}} =\diag{\R_{km},\cdots,\R_{km+m-1}},\eqnlabel{Rk}
\end{align}
where 
\begin{align*}
\Q_{km+j} &:= \Exp{\w_{km+j}\w^T_{km+j}},\\
\R_{km+j} &:= \Exp{\n_{km+j}\n^T_{km+j}}.
\end{align*}

\subsection{Uncertainty Propagation and Measurement Update}
The prior statistics of $\X_k$ are related to the posterior statistics of $\x_{km}$ i.e. $(\mub_{km}^+,\Sigb^+_{km})$ as 
\begin{align}
\Xbar_k^- & := \Exp{\X_k} = \sA_k\mub_{km}^+,\\
\nonumber \vo{P}^-_k & := \Exp{(\X_k^- - \Xbar_k^-)(\X_k^- - \Xbar_k^-)^T},\\
& = \sA_k\Sigb^+_{km}\sA_k^T + \sB_k\sQ_k\sB_k^T.
\end{align}
Prior statistics of the augmented state, i.e. $\Xbar_k^-$ and $\vo{P}^-_k$, can be updated using the augmented measurements $\Y_k$ to obtain posterior $(\Xbar_k^+,\vo{P}^+_k)$, using similar steps as in standard Kalman filtering, i.e.

\begin{align}
\Xbar_k^+ &:= \sA_k\mupp_{km} + \sK_k(\Y_k-\sC_k\sA_k\mupp_{km}),\\
\vo{P}^+_k &:= (\mathbf{I}-\sK_k\sC_k)\vo{P}^-_k, \eqnlabel{P+}
\end{align} 
where
\begin{align}
\sK_k := \vo{P}^-_k\sC_k^T\Big[\sC_k\vo{P}^-_k\sC_k^T+\sR_k\Big]^{-1}. 
\eqnlabel{KalmanGain}
\end{align}
The state at time $t_{(k+1)m}$ can be determined from $\X_k$ as
$$\x_{(k+1)m} := \vo{M}_m\X_k,$$ where
\begin{align}
\vo{M}_m := \begin{bmatrix} \vo{0}_{n_x\times n_x(m-1)} & \I_{n_x}\end{bmatrix}\eqnlabel{Mn}.
\end{align}

The posterior statistics of $\x_{(k+1)m}$ can then be determined from the posterior statistics of $\X_k$ using
\begin{align}
\mub_{(k+1)m}^+ := \vo{M}_m\Xbar_k^+, \text{ and } \Sigb_{(k+1)m}^+ :=\vo{M}_m \vo{P}^+_k \vo{M}_m^T.
\end{align}
  
\subsection{Optimal Sensor Precision for a Single Measurement Update}
Here we present a convex optimization framework for determining the nosiest sensors, for which the estimation errors are below a given upper bound after \textit{one measurement update}. That is, maximize $\trace{\sR_k}$ or minimize $\trace{\sS_k}$ where $\sS_k := \sR_k^{-1}$, for which $\trace{\Sigb_{(k+1)m}^+} \le \gamma_d$, given $\Sigb_{km}^+$. This is achieved by solving the following optimization problem.

\begin{theorem}\label{thm:1}
Optimal sensor precision $\vo{s}_k \in\real^{n_{y_{k,m}}}\ge 0$, which satisfies $\trace{\Sigb_{(k+1)m}^+}\le\gamma_d$, is given by the solution of the following optimization problem,
\begin{equation}\left.
\begin{aligned}
&\min_{\vo{s}_k,\sK_k,\vo{F}}{\trace{\vo{W}\sS_k}}, \text{ subject to }\\
&\begin{bmatrix}
\vo{F} & \M_{12}\sqrt{\vo{P}^-_k} & \sK_k\\
(\ast)^T & \I_{mn_x} & \vo{0}_{mn_x \times n_{y_{k,m}}}\\
(\ast)^T & (\ast)^T & \sS_k
\end{bmatrix} \ge 0\\
& 0 \le \vo{s}_k \le \vo{s}_k^\text{max},
\end{aligned} \right \}
\eqnlabel{thm1}
\end{equation}
where $\M_{12} := \M_m(\I_{mn_x}-\sK_k\sC_k)$ , $\sK_k$ and $\vo{F}$ are the design variables with $\trace{\vo{F}} \le \gamma_d$, $\M_n$ and system parameter $\sC_k$ is defined in \eqn{Mn} and \eqn{Ck}. The variable $\vo{W}$ is a diagonal matrix, which is user defined and serves as a normalizing weight on $\sS_k:= \sR_k^{-1}$, where $\sR_k$ is defined in \eqn{Rk}.
\end{theorem}

\begin{\pf}
We can write the posterior error variance as
\begin{align}
\vo{P}^+_k = (\I_{mn_x}-\sK_k\sC_k)\vo{P}^-_k(\I_{mn_x}-\sK_k\sC_k)^T + \sK_k\sR_k\sK_k^T.
\eqnlabel{posterior:PK}
\end{align}
The optimal $\sK_k$ is determined by minimizing $\trace{\vo{P}^+_k}$ and is given by \eqn{KalmanGain}. However, in this formulation, we leave $\sK_k\in\real^{mn_x\times n_{y_{k,m}}}$ as a variable, and write $\trace{\M_m\vo{P}^+_k \M_m^T}\le \gamma_d$ equivalently as
\begin{align*}
& \vo{F} - \M_{12}\vo{P}^-_k\M_{12}^T - \M_m\sK_k^T\sR_k\sK_k^T\M^T_m \ge 0,\\
\text{ and } & \trace{\vo{F}} \le \gamma_d,
\end{align*}
where $\M_{12} := \M_m(\I_{mn_x}-\sK_k\sC_k)$ and $\vo{F} \in \mathbb{S}^{n_x}_+$. Representing $\sqrt{\vo{P}^-_k}$ as the principal matrix square-root of $\vo{P}^-_k$, substituting $\sS_k:=\sR_k^{-1}$ which is defined in \eqn{Rk} , and using Schur complement we get the following LMI \cite{boyd1994linear},
\begin{align*}
\begin{bmatrix}
\vo{F} & \M_{12}\sqrt{\vo{P}^-_k} & \M_m\sK_k\\
(\ast)^T & \I_{mn_x} & \vo{0}_{mn_x \times n_{y_{k,m}}}\\
(\ast)^T & (\ast)^T & \sS_k
\end{bmatrix} \ge 0.
\end{align*}
Combining the inequalities and minimizing $\trace{\vo{W}\sS_k}$ we get \eqn{thm1}.
\end{\pf}

%
%
%

\begin{remark}\label{rem:improveSparseness}
In theorem \ref{thm:1} the sparseness of the solution can be improved by iteratively solving the optimization problem \eqn{thm1} with weights $\vo{W}_{j+1} := (\sS_{k,j}^\ast + \epsilon\vo{I})^{-1}$, with $\vo{W}_{1}:=\I_{n_{y_{k,m}}}$ and $\vo{I}$ is the identity matrix of appropriate dimension, where subscript $j$ denotes the iteration index \cite{rao1999affine, candes2008enhancing} and $\sS_{k,j}^\ast$ is the solution to \eqn{thm1} in the $j^{\text{th}}$ iteration of improving sparsity in the solution.
\end{remark}

\section{Optimal Sensor Precision for Bounded Steady-State Errors}\label{sec:steady-state}
In this section, we present the result that determines the optimal sensor precision for bounded steady-state error, assuming the system to be $m$-periodic. If the system in \eqn{processDynamics} is $m$-periodic, i.e. $\A_{km+j} = \A_{(k+1)m+j}$, $\B_{km+j} = \B_{(k+1)m+j}$, and $\C_{km+j} = \C_{(k+1)m+j}$ for $j=1,\cdots,m$; it will be of interest to determine the sensing precisions that bound the steady-state errors, assuming it exists.

\subsection{Augmented Dynamical System}
From \eqn{augDyn}, the augmented dynamics of the $m$-periodic system is given by,
\begin{align}
\x_{(k+1)m} & = \M_m\sA_k\x_{km} + \M_m\sB_k\W_{k}.\eqnlabel{augdyn}
\end{align}
In this section we generalize the sensor model in \eqn{sensing} by including the process noise in the measurement\cite{Deshpande_2017}. This scenario, for example, occurs in measurements from accelerometers where the disturbance forces algebraically impact the accelerations. The new measurement model is therefore,
\begin{equation}
\y_{k} = \C_k\x_k + \D_k\w_k + \n_{k}. \eqnlabel{sensing:D}
\end{equation}
Consequently, the augmented sensor model is 
\begin{align}
\Y_{k} &= \sC_k\X_k + \sD_k\W_k + \N_{k}, \nonumber \\
& = \sC_k(\sA_k\x_{km} + \sB_k\W_{k}) + \sD_k\W_k + \N_{k},\nonumber\\
& = \sC_k\sA_k\x_{km} + \left(\sC_k\sB_k + \sD_k\right)\W_k + \N_{k},\eqnlabel{augmeas}
\end{align}
where $\sD_k := \diag{\D_{km+1},...,\D_{(k+1)m}}$.

The presence of $\W_k$ in \eqn{augmeas} makes derivation of the Kalman filter complicated. This is circumvented by assuming the process noise to be colored, or filtered white noise. That is, we model the process noise as
\begin{align}
\Z_{k+1} = \sG\Z_k + \sH\vo{\Lambda}_k, \; \W_k = \Z_k,\eqnlabel{coloredNoise}
\end{align}
where $\vo{\Lambda}_k$ is white noise, $\Z_k$ is the filter state, and the pair $(\sG_k,\sH_k)$ defines the filter. 

If white noise $\vo{\lambda}_k\in\real^{n_w}$ is filtered via 
\begin{equation}
\z_{km+1} = \G\z_{km} + \H\vo{\lambda}_{km}, \eqnlabel{filter}
\end{equation}
then for the augmented system,
\begin{align}
\Z_k := \begin{pmatrix}\z_{km} \\ \vdots \\ \z_{(k+1)m-1}\end{pmatrix},\; \vo{\Lambda}_k := \begin{pmatrix}\vo{\lambda}_{km} \\ \vdots \\ \vo{\lambda}_{(k+1)m-1}\end{pmatrix},
\end{align}
$\sG := \I_{q}\otimes\G$, and $\sH := \I_{q}\otimes\H$, where $\G\in\real^{n_w\times n_w}$ and $\H\in\real^{n_w\times n_w}$ define the filter in \eqn{filter}.

Introducing a new state variable
\begin{align}
\Gamb_k := \begin{bmatrix} \x_{km} \\ \Z_k \end{bmatrix} \in \real^{N_x},
\end{align}
where $N_x := n_x+mn_w$, we can write the dynamics of $\Gamb_k$ and measurement $\Y_k$ as
\begin{subequations}
\begin{align}
\Gamb_{k+1} = \sA_m\Gamb_k + \sB_m\vo{\Lambda}_k, \eqnlabel{gamDyn}\\
\Y_k = \sC_m\Gamb_{k} + \N_k. \eqnlabel{gamY}
\end{align}
\end{subequations}
where 
\begin{subequations}
\begin{align}
\sA_m &:= \begin{bmatrix} \M_m\sA_k & \M_m\sB_k \\ \vo{0}_{qn_w\times n_x} & \sG \end{bmatrix},\\
\sB_m &:= \begin{bmatrix}\vo{0}_{n_x \times qn_w} \\ \sH\end{bmatrix},\\
\sC_m &:= \begin{bmatrix}\sC_k\sA_k & (\sC_k\sB_k + \sD_k)\end{bmatrix}.
\end{align}
\end{subequations}
Note that for the $m$-periodic system, the matrices $\sA_m$, $\sB_m$, and $\sC_m$ are time invariant. 

States $\x_{km}$ can be recovered from $\Gamb_k$ as
\begin{align}
\x_{km} = \M_x\Gamb_k,
\end{align}
where $\M_x$ is a mask-matrix defined by
\begin{align}
\M_x := \begin{bmatrix} \I_{n_x} & \vo{0}_{n_x \times qn_w}\end{bmatrix}\eqnlabel{Mx}.
\end{align}
We next define 
\begin{align}
\sQ_m := \Exp{\vo{\Lambda}_k\vo{\Lambda}_k^T}, \text{ and }  \sR_m := \Exp{\N_k\N^T_k}.
\end{align}
For steady-state analysis, we assume $(\sA_m,\sR_m^{1/2}\sC_m)$ is detectable and $(\sA_m,(\sB_m\sQ_m\sB^T_m)^{1/2})$ is stabilizable \cite{anderson2012optimal}.

\subsection{Steady-state Variance}
Let $\bar{\Gamb}^-_{k}$ and $\P^-_k$ be the prior mean and variance of $\Gamb_{k}$ at time $k$. This defines the prior random variable $\Gamb_{k}^- \sim \mathcal{N}(\bar{\Gamb}^-_{k},\P^-_{k})$, where $\mathcal{N}(\cdot,\cdot)$ defines a Gaussian distribution.

In Kalman filtering we assume the posterior is a linear function of the prior and the measurement, i.e. 
\begin{align}
\Gamb^+_{k} := (\I_{N_x} - \sK_k\sC_m)\Gamb^-_{k} + \sK_k\Y_{k}, \eqnlabel{postVar}
\end{align}
where $\sK_k\in\real^{N_x\times n_{y_{k,m}}}$ is the unknown gain.

The coefficient $\sK_k$ is determined by minimizing the posterior variance. However, in this formulation, we leave it as a free variable along with $\sR_m$. Both these variables will be jointly determined in a single optimization problem, presented in theorem \ref{thm:2}.

Using \eqn{postVar}, the posterior variance is given by
\begin{align}
\P^+_{k} = (\I_{N_x} - \sK_k\sC_m)\P^-_{k}(\I_{N_x} - \sK_k\sC_m)^T + \sK_k\sR_m\sK^T_k. \eqnlabel{ricc1}
\end{align}

Using \eqn{gamDyn}, the prior mean and variance of $\Gamb_{k}$ at time $k+1$ is given by
\begin{subequations}
\begin{align}
\bar{\Gamb}^-_{k+1} &= \sA_m\bar{\Gamb}^+_{k},\\
\P^-_{k+1} &= \sA_m\P^+_k\sA^T_m + \sB_m\sQ_m\sB_m^T, \eqnlabel{propVar}
\end{align}
\end{subequations}
which defines the random variable $\Gamb_{k+1}^- \sim \mathcal{N}(\bar{\Gamb}^-_{k+1},\P^-_{k+1})$.

Replacing $\P^-_{k}$ from \eqn{ricc1} in \eqn{propVar}, we get the propagation equation for the prior variance
\begin{align}
\P^-_{k+1} =& \sA_m(\I_{N_x} - \sK_k\sC_m)\P^-_{k}(\I_{N_x} - \sK_k\sC_m)^T\sA^T_m \nonumber\\ &+ \sA_m\sK_k\sR_m\sK^T_k\sA^T_m + \sB_m\sQ_m\sB_m^T. \eqnlabel{RDE}
\end{align}
Steady-state variance $\P_\infty$ is determined by solving 
\begin{align}
\P_\infty =& \sA_m(\I_{N_x} - \sK_\infty\sC_m)\P_\infty(\I_{N_x} - \sK_\infty\sC_m)^T\sA^T_m  \nonumber\\&+ \sA_m\sK_\infty\sR_m\sK^T_\infty\sA^T_m + \sB_m\sQ_m\sB_m^T, \eqnlabel{ARE}
\end{align}
where $\sK_\infty$ is the steady-state gain. The steady-state variance of $\x_{(k+1)m}$ is then given by $\Sigb_\infty := \M_x\P_\infty \M_x^T$.

\begin{remark}
Equations \eqn{RDE} becomes the Riccati difference equation (RDE) if $\sK_k$ is determined by minimizing the posterior variance. Consequently, it transforms \eqn{ARE} to the algebraic Riccati equation (ARE). That is, for $\sK_k$ given by \eqn{KalmanGain},
\eqn{RDE} transforms to
\begin{align}
\vo{P}_{k+1}  =& \sA_m(\vo{P}_{k}-\vo{P}_{k}\sC_m^T[\sC_m \vo{P}_{k}\sC_m^T+\sR_m]^{-1}\sC_m \vo{P}_{k})\sA_m^T \nonumber\\ &+\sB_m\sQ_m\sB^T_m,\eqnlabel{RDE1}
\end{align} and
\eqn{ARE} transforms to
\begin{align}
\vo{P}_\infty  =&\sA_m(\vo{P}_{\infty}-\vo{P}_\infty\sC_m^T[\sC_m \vo{P}_{\infty}\sC_m^T+\sR_m]^{-1}\sC_m \vo{P}_{\infty})\nonumber \\ &\times\sA_m^T  +\sB_m\sQ_m\sB^T_m,\eqnlabel{ARE1}
\end{align}
Equation \eqn{ARE1} has a unique positive semi-definite solution if $(\sA_m,\sR_m^{1/2}\sC_m)$ is detectable and $(\sA_m,(\sB_m\sQ_m\sB^T_m)^{1/2})$ is stabilizable.
\end{remark}

\subsection{Optimal Sensor Precision for Bounded Steady-State Estimation Error}
Let $\vo{P}^d_\infty$ be the desired steady-state error, and let us assume that it is the solution of \eqn{ARE1} for some $\sR^d_m$, i.e. 
\begin{align*}
\vo{P}^d_\infty =&  \sA_m(\vo{P}^d_{\infty}-\vo{P}^d_\infty\sC_m^T\left[\sC_m \vo{P}^d_{\infty}\sC_m^T +\sR^d_m\right]^{-1}\sC_m \vo{P}^d_{\infty})\nonumber\\ &\times\sA_m^T  + \sB_m\sQ_m\sB^T_m.
\end{align*}
Therefore, for any $\sR_m - \sR^d_m \le 0$
\begin{align}
\vo{P}^d_\infty \ge&  \sA_m(\vo{P}^d_{\infty}-\vo{P}^d_\infty\sC_m^T[\sC_m \vo{P}^d_{\infty}\sC_m^T+\sR_m]^{-1}\sC_m \vo{P}^d_{\infty})\nonumber\\ &\times\sA_m^T  +\sB_m\sQ_m\sB^T_m,\eqnlabel{ARE-relaxed}
\end{align}
which makes the solution of the RDE monotonic.

According to Lemma 2 in \cite{bitmead1985monotonicity}, if for some $k$ the solution of the RDE in \eqn{RDE1} is monotonic, i.e. $\vo{P}_{k} \ge \vo{P}_{k+1}$, then $\vo{P}_{k+i} \ge \vo{P}_{k+i+1}$ for all $i \ge 1$. Therefore, \eqn{ARE-relaxed}, guarantees that the evolution of $\vo{P}^d_\infty$ is monotonic, and $\vo{P}_\infty$ is guaranteed to satisfy $\vo{P}^d_\infty \ge \vo{P}_\infty$. The gap in the inequality can be minimized by maximizing $\trace{\sR_m}$. 

The optimization problem that determines the optimal sensor precision to guarantee $\trace{\vo{M}_x\P_\infty^d\vo{M}_x^T}\le\gamma_d$, for a given $\gamma_d$ is presented as the following theorem.

\begin{theorem} \label{thm:2}
Optimal sensor precision $\vo{s} \in\real^{n_{y_{k,m}}}\ge 0$, which satisfies $\trace{\vo{M}_x\P_\infty^d\vo{M}_x^T}\le\gamma_d$ is given by the solution of the following optimization problem,
\begin{subequations}
\small
\begin{align}
& \min_{\vo{s}_m,\Z,\vo{P}^d_\infty,\sK_\infty}{\trace{\W\sS_m}} \text{ subject to }\\
&\begin{bmatrix}
\M_{11} & \M_x\sA_m(\I_{N_x} - \sK_\infty\sC_m) & \M_x\sA_m\sK_\infty \\
(\ast)^T & \Z & \vo{0}_{N_x\times n_{y_{k,m}}}\\
(\ast)^T & (\ast)^T & \sS_m
\end{bmatrix} \ge 0,\\
&\begin{bmatrix}
2\I_{N_x} & \P^d_{\infty} & \Z\\
\P^d_{\infty} & \frac{1}{\delta}\I_{N_x} & \vo{0}_{N_x\times N_x}\\
\Z & \vo{0}_{N_x\times N_x} & \delta\I_{N_x}
\end{bmatrix} \ge 0, \eqnlabel{delta}\\
&\trace{\vo{M}_x\P_\infty^d\vo{M}_x^T}\le\gamma_d,\\
& 0 \le \vo{s}_m \le \vo{s}_\text{max},
\end{align}
\eqnlabel{thm2}
\end{subequations}\normalsize
$\M_{11} := \M_x\left(\P^d_{\infty} - \sB_m\sQ_m\sB_m^T\right)\M^T_x$, $\M_x$ defined in \eqn{Mx}, $\sB_m$ and $\sQ_m$ are system parameters, $\sS_m := \diag{\vo{s}_m}$, $n_{s}:=\sum_{j=1}^m p_{j}$, $\Z\in\mathbb{S}_+^{N_x}$ is a design variable, and $p_j$ is the dimension of the j$^\text{th}$ sensor. Variables $\gamma_d > 0$ and $\delta > 0$ are user specified. The variable $\vo{W}$ is a diagonal matrix, which is also user defined, and serves as a normalizing weight on $\sS_m$.
\end{theorem}

\begin{\pf}
From \eqn{RDE}, monotonicity of $\P_\infty^d$ is guaranteed if
\begin{align*}
\P^d_{\infty} \ge& \sA_m(\I_{N_x} - \sK_\infty\sC_m)\P^d_{\infty}(\I_{N_x} - \sK_\infty\sC_m)^T\sA^T_m \nonumber \\ &+ \sA_m\sK_\infty\sR_m\sK^T_\infty\sA^T_m + \sB_m\sQ_m\sB_m^T.
\end{align*}
Introducing a new variable $\Z\in\mathbb{S}_+^{N_x}$, and the relaxation $$\Z^{-1}\ge\P^d_{\infty},$$ the condition for monotonicity can then be written as 
\begin{align*}
\P^d_{\infty} \ge& \sA_m(\I_{N_x} - \sK_\infty\sC_m)\Z^{-1}(\I_{N_x} - \sK_\infty\sC_m)^T\sA^T_m  \nonumber \\ &+ \sA_m\sK_\infty\sR_m\sK^T_\infty\sA^T_m + \sB_m\sQ_m\sB_m^T.
\end{align*}
However, we want to enforce monotonicity of $\M_x\P^d_{\infty}\M_x^T$, i.e.
\begin{align*}
\M_x\P^d_{\infty}\M^T_x \ge& \M_x\sA_m(\I_{N_x} - \M_x\sK_\infty\sC_m)\Z^{-1}\nonumber\\ &\times(\I_{N_x} - \M_x\sK_\infty\sC_m)^T\sA^T_m\M^T_x \\ &+ \M_x\sA_m\sK_\infty\sR_m\sK^T_\infty\sA^T_m\M^T_x \nonumber\\&+ \M_x\sB_m\sQ_m\sB_m^T\M^T_x.
\end{align*}
Using Schur complement \cite{caverly2021lmi}, and substituting $\sS_m := \sR_m^{-1}$, we get 
$$
\begin{bmatrix}
\M_{11} & \M_x\sA_m(\I_{N_x} - \sK_\infty\sC_m) & \M_x\sA_m\sK_\infty \\
(\ast)^T & \Z & \vo{0}_{N_x\times n_{y_{k,m}}}\\
(\ast)^T & (\ast)^T & \sS_m
\end{bmatrix} \ge 0,
$$
where $\M_{11} := \M_x\left(\P^d_{\infty} - \sB_m\sQ_m\sB_m^T\right)\M^T_x$.

The relaxation $\Z^{-1}\ge\P^d_{\infty}$ can be written as $\P^d_{\infty}\Z \le \I_{N_x}$, which is non convex. However, we know that
$$ \P^d_{\infty}\Z + \Z\P^d_{\infty} \le \delta\P^d_{\infty}\P^d_{\infty} + \frac{1}{\delta}\Z\Z,$$
for a given $\delta$ following from a special case of Young's relation \cite{caverly2021lmi}. Therefore, 
\begin{equation}
\delta\P^d_{\infty}\P^d_{\infty} + \frac{1}{\delta}\Z\Z \le 2\I_{N_x}, \eqnlabel{young}
\end{equation}
guarantees $\P^d_{\infty}\Z \le \I_{N_x}$. The inequality in \eqn{young}, can be written as the following linear matrix inequality
$$
\begin{bmatrix}
2\I_{N_x} & \P^d_{\infty} & \Z\\
\P^d_{\infty} & \frac{1}{\delta}\I_{N_x} & \vo{0}_{N_x\times N_x}\\
\Z & \vo{0}_{N_x\times N_x} & \delta\I_{N_x}
\end{bmatrix} \ge 0.
$$
The optimal precision is given by minimizing $\trace{\sS_m}$.
\end{\pf}

\begin{remark}
The parameter $\delta$ can be tweaked to improve the solution, using techniques from successive convex over-bounding techniques described in \cite{warner2017iterative}.
\end{remark}

\begin{remark}
Like in theorem \ref{thm:1}, the sparseness of the solution can be improved by iteratively solving the optimization problem in theorems \ref{thm:2} with weights $\vo{W}_{j+1} := (\sS_m^\ast)^{-1}_j$, with $\vo{W}_{1}:=\I_{n_{s}}$, where subscript $j$ denotes the iteration index and $(\sS_m^\ast)_j$ denotes the optimal $\sS_m$ calculated at the $j^{\text{th}}$ iteration.
\end{remark}

\begin{remark}\label{rem:conservative}
The optimal $\sS^\ast_m$ can be conservative. The actual steady-state variance, denoted by $\vo{P}_{\infty}$, can be much smaller than $\vo{P}^d_{\infty}$. At the same time, it is possible that $\vo{P}_{\infty}$ is smaller than $\vo{P}^d_{\infty}$ without requiring $\vo{P}^d_{\infty}$ to be monotonic. Thus, theorem \ref{thm:2} is conservative and this results in more precision than that required to achieve $\vo{P}^d_{\infty}$. 

We remove the conservativeness by applying the following strategy. The optimization problem in theorem \ref{thm:2} is solved to determine the sparse solution $\sS^\ast_m$ for which $\vo{P}^d_{\infty}$ is contractive. It is then scaled by $\xi$ to reduced the gap between $\vo{P}^d_{\infty}$ and $\vo{P}_{\infty}$, where optimal $\xi^\ast$ is obtained using the bisection algorithm described in Algorithm \ref{alg:scaling}.

\begin{algorithm}
 \caption{Bisection algorithm for optimal scaling of sensor precision.}
 \label{alg:scaling}
 \begin{algorithmic}
 \item Define: $\xi_\text{min} := 0$
 \item Define: $\xi_\text{max} := 10^3$ \texttt{\# Something large}
 \item Solve optimization problem \eqn{thm2} to get $\sR^\ast_m := (\sS^\ast_m)^{-1}$.
	\item Define: \texttt{MAXITER} = $100$ \texttt{\# Something large}
 \item {\ttfamily \bfseries for} {\tt i = 1:MAXITER}
	\item \hspace{0.5cm} $\xi = \frac{1}{2}(\xi_\text{min} + \xi_\text{max})$
	\item \hspace{0.5cm} $\vo{P}_\text{ss} :=$ solution of ARE in \eqn{ARE} with noise $\xi\sR_m^\ast$
	\item \hspace{0.5cm}{\ttfamily \bfseries if} $\trace{\vo{M}_x\vo{P}_\text{ss}\vo{M}^T_x} < \gamma_d$\\
		 \hspace{1cm} $\xi_\text{min} := \xi$ \\
		 \hspace{0.5cm} {\ttfamily \bfseries else} \\ 
		 	\hspace{1cm} $\xi_\text{max} := \xi$ \\
			\hspace{0.5cm} {\ttfamily \bfseries end}\\
	{\ttfamily \bfseries end}
 \end{algorithmic}
\end{algorithm}
\end{remark}

\section{Examples}\label{sec:examples}
Next, we apply theorems 1 and 2 to three estimation problems related to aerospace engineering, highlighting their engineering value. The first example demonstrates the application of theorem 2 to determine redundant sensing architectures for linear time invariant (LTI) systems. The second example demonstrates accurate state estimation with low-cost sensors. Finally, the third example demonstrates sensor scheduling for a linear time-varying system.

\subsection{Time-Invariant System: Flight Control Example}
Here we demonstrate practical applications of the result presented in theorem \ref{thm:2} to a linear \textit{time-invariant} discrete-time system. It is applied to a steady-state estimation problem for an aircraft model. We first present the details of the aircraft model. We then present two examples, which highlight different applications of theorem \ref{thm:2}.

\subsubsection{Model}\label{sec:model}
Let us consider the longitudinal motion model of an aircraft, where the states of the system are velocity $V$ in $ft/s$, angle of attack $\alpha$ in $rad$, pitch angle $\theta$ in $rad$, and pitch rate $q$ in $rad/s$, i.e.
\begin{equation}
\x := \begin{bmatrix} V & \alpha  & \theta  & q\end{bmatrix}^T.
\end{equation}
We consider onboard sensors that measure body acceleration $\dot{u}\,(ft/s^2)$ along roll axis, body acceleration $\dot{w}\,(ft/s^2)$ along yaw axis, angle of attack $\alpha(rad)$, pitch rate $q(rad/s)$, and dynamic pressure $\bar{q}:=\frac{1}{2}\rho V^2\, (lb/ft^2)$, where $\rho$ is the atmospheric density. Variables $u$ and $w$ are defined as $u:=V\cos(\alpha)$, and $w:=V\sin(\alpha)$.

Therefore, the vector of measured outputs is  
\begin{equation}
\y:= \left[\dot{u},\dot{w},\alpha,q,\bar{q}\right]^T. \eqnlabel{sensors} 
\end{equation} 
In a typical aircraft, these measurements are available from the accelerometers, angle-of-attack sensors, gyro sensors, and pitot tube, respectively.

The dynamics and measurement model is given by the following equations
\begin{subequations}
\begin{align}
\frac{d\x}{dt} &= \A\x + \vo{B} d,\\
\vo{y} &= \C\x + \vo{D} d + \vo{n},
\end{align}\eqnlabel{dynamics}
\end{subequations}
where \footnote{More accurate data is available upon request.} 
\begin{align*}
&\A = \begin{bmatrix} 
 -0.0179 &33.2244 &-32.1700 &0.6728\\
 -0.0001 &-1.4528 &  0 &0.9323\\
   0 &  0 &  0 &1.0000\\
 -0.0000 &-4.1970   &0 &-1.8836
       \end{bmatrix},\\ 
&\B = \begin{bmatrix} 
  0.5697& -0.0029 & 0 & -0.4670
 \end{bmatrix}^T, \\    
&\C = 10^{3}\times\begin{bmatrix}
  -0.0000 &0.0332 &-0.0322 &0.0007\\
  -0.0001 &-1.3544 &  0 &0.8692\\
   0  &0.0010 &  0 &  0\\
   0  &  0 &  0 &0.0010\\
 0.0017  & 0 &  0 &  0
  \end{bmatrix}, \\
&\D = \begin{bmatrix} 0.5697 & -2.7345 & 0 & 0 & 0 \end{bmatrix}^T,\\
\end{align*}
$d$ is the disturbance acting on the system, and $\vo{n}$ is the sensor noise. Note that there is a direct feed-through term because the disturbance directly impacts the acceleration measurements. In this example, we model the disturbance as a filtered white noise, filtered by $\frac{1}{s/\omega_c+1}$, where $\omega_c$ is the cutoff frequency. The disturbance is generated by a vibrating control surface in the aircraft.

Let the filter state be $x_d$, and the filter dynamics be given by 
$$
\dot{x}_d = \omega_c(-x_d + w), \text{ and } d = x_d,
$$
where $w$ is white noise with a given variance. In this example, we choose the variance to be $5$ $\text{deg}^2$. This corresponds to small angular deflections in the control surface, which is measured in degrees.

\subsubsection{\textbf{Example 1:} Redundancy in Sensing}
Here we discretize the continuous time system with sampling time $\Delta t:=0.01 \ s $ and determine the least precision needed to achieve a steady-error that satisfies $\trace{\M_x\P_\infty\M_x^T}\le\gamma_d$ for $\gamma_d:=0.1$, where $\P_\infty$ quantifies the actual steady-state error. In this example, theorem \ref{thm:2} is applied with $m = 1$. For the filter, the cutoff frequency $\omega_c$ is chosen to be $10$ rad/s. Finally, $\delta = 200$ was chosen to implement the constraint in \eqn{delta}.

\begin{figure}[h!] \centering
\includegraphics[width=0.4\textwidth]{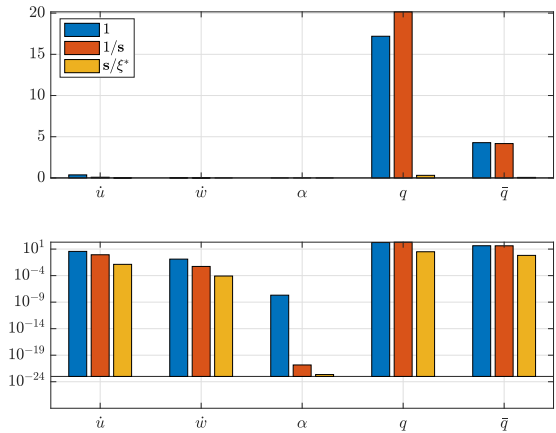}
\caption{Sensor precisions for the five sensors from different algorithms. The respective units are $(ft/s^2)^2$, $(ft/s^2)^2$, $rad^2$, $(rad/s)^2$, and $(lb/ft^2)^2$}
\figlabel{sr_all}
\end{figure}

\Fig{sr_all} shows the sensor precisions from the unweighted optimization (indicated by legend ``$1$''), the precisions from iteratively weighted optimization to improve sparseness (indicated by legend ``$1/\textbf{s}$''), and finally, the scaled precision to remove the conservativeness in the optimal solution (indicated by legend ``$\textbf{s}/\xi^\ast$''). The top-panel in \fig{sr_all} shows the sparse solution, and the bottom panel shows the same data on the logarithmic scale. We observe that out of the five sensors chosen in the design, only two significantly contribute to the required estimation accuracy. 

Iteratively weighted optimization significantly improves the sparseness in the solution by several orders of magnitude. We also observe that the weighted optimization solution is conservative, and the precisions can be further reduced to get closer to the boundary of $\trace{\M_x\P_\infty\M_x^T}\le\gamma_d$. 

\Fig{sr} shows the scaled optimal precisions for $\xi^\ast = 64.1106$. The precision values in \fig{sr} indicate that only angular velocity measurement $q$ and dynamic pressure data $\bar{q}$ are needed in higher precision to estimate all four states of the system with the required accuracy. 

\begin{figure}[h!] \centering
\begin{subfigure}[t]{0.45\textwidth}\centering
\includegraphics[width=\textwidth]{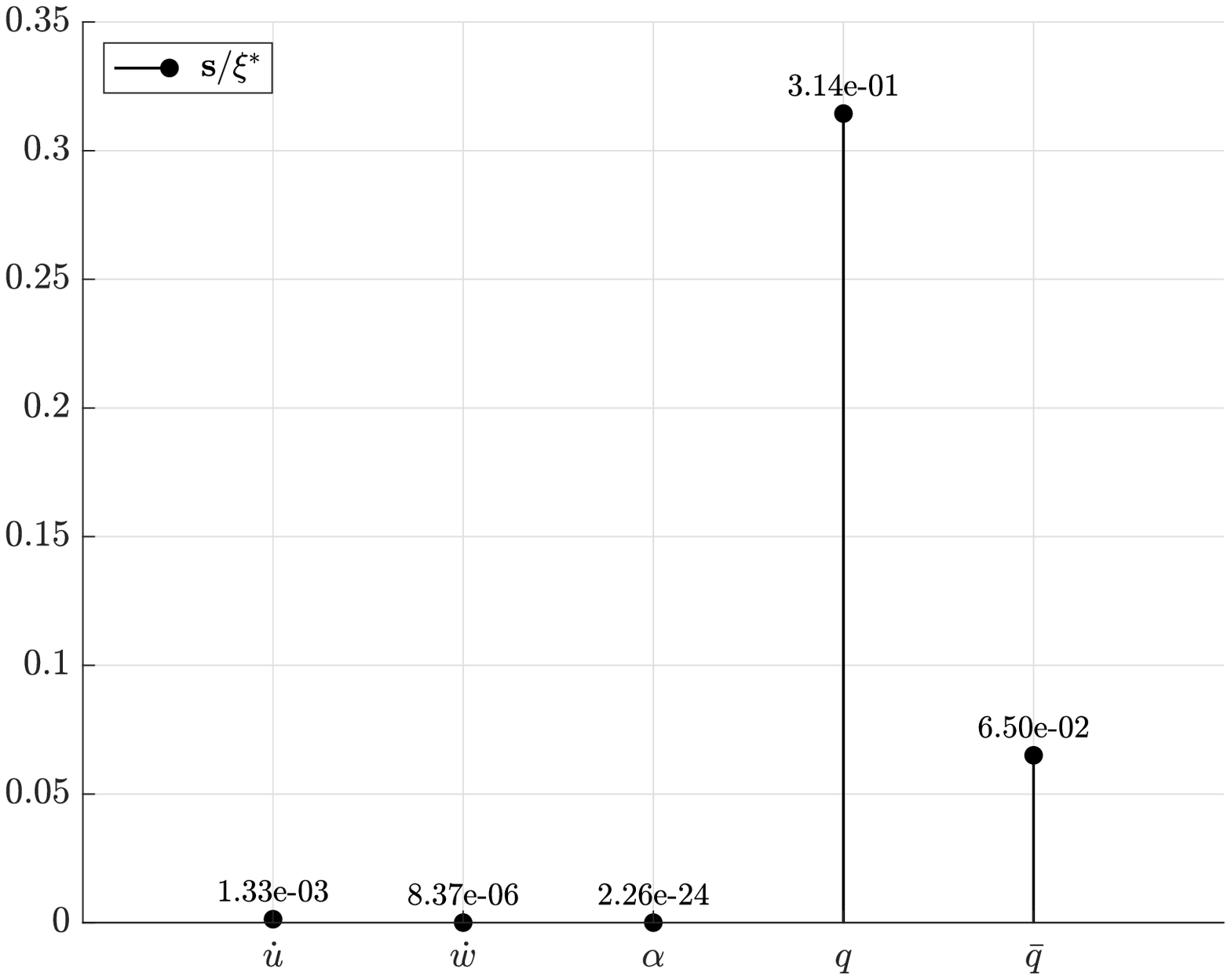}
\caption{Optimal scaled sensor precisions satisfying $\trace{\M_x\P_\infty\M_x^T}\le\gamma_d$ for $\gamma_d:=0.1$.}
\figlabel{sr}
\end{subfigure}\hfill
\begin{subfigure}[t]{0.45\textwidth}\centering
\includegraphics[width=\textwidth]{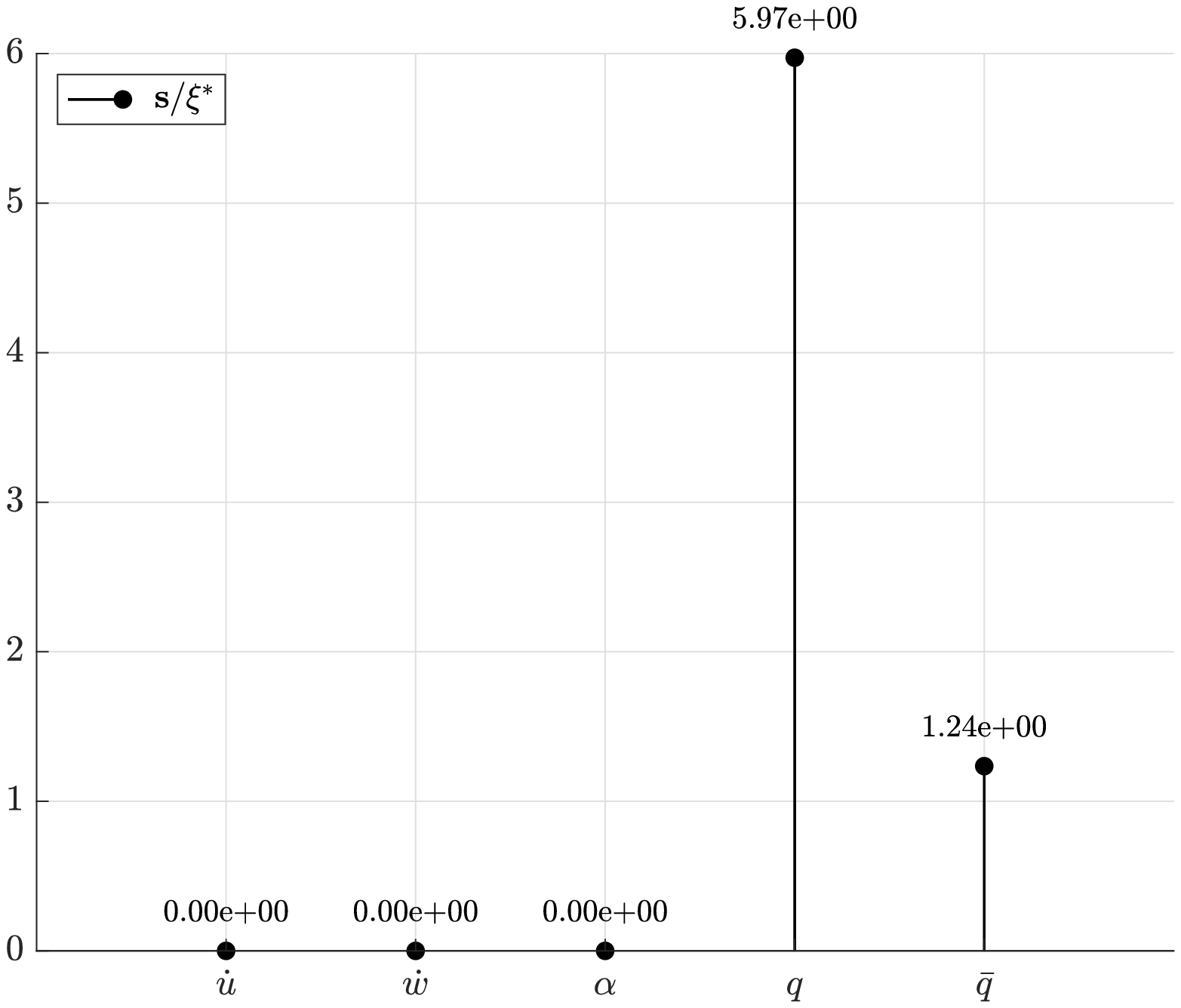}
\caption{State estimation with only $q$ and $\bar{q}$: required precisions to achieve $\trace{\M_x\P_\infty\M_x^T}\le\gamma_d$ for $\gamma_d:=0.1$.}
\figlabel{sr_moresparse}
\end{subfigure}
\end{figure}

To completely remove the sensor for $\dot{u},\dot{v}$, and $\alpha$, we can set their corresponding precisions to exactly zero, prior to $\xi$ scaling. Since detectability and stabilizability conditions are verified to be satisfied for this sensor configuration, \eqn{ARE} has a unique solution. Therefore, using algorithm \ref{alg:scaling} we can determine the optimal $\xi$ scaling that guarantees $\trace{\M_x\P_\infty\M_x^T}\le\gamma_d$. For this example, we get $\xi^\ast = 3.375$, and the scaled sensor precisions are shown in \fig{sr_moresparse}. We see from \fig{sr_moresparse} that the required precisions for $q$ and $\bar{q}$ in this case are much higher than those in \fig{sr}. 



Therefore, from a sensor pruning perspective, an ad-hoc approach would be to start with a dictionary of sensors and determine the optimal sensor precisions using theorem \ref{thm:2}, then assign zero precisions to those sensors with small precisions, and finally apply algorithm \ref{alg:scaling} to arrive at the optimal precisions of the reduced number of sensors. More sophisticated algorithms \cite{Dhingra_2014, Lin_2013, Tzoumas_2016} for sensor pruning or sensor selection can also be applied. These algorithms assume sensor precisions are known, which can be determined from theorems 1 or 2 .

\subsubsection{\textbf{Example 2:} Accurate State Estimation with Low Precision Sensors}
In this section, we apply theorem \ref{thm:2} to explore the tradeoff between sensing rate and sensing precision. We use the same F16 example described in \S\ref{sec:model}. In this example, the continuous-time model is discretized with $dt = 1/1000 \ s$, using Tustin's method \cite{franklin1998digital}. The augmented system is created with pitch rate $q=10 (rad/s)$. This formulation captures a scenario where the sensor data is available at $1$ Khz, but the state estimates are needed at $100$ Hz. We assume that the state's estimates are used by some control law executing at $100$ Hz.

\begin{figure}[h!] \centering
\includegraphics[width=0.35\textwidth]{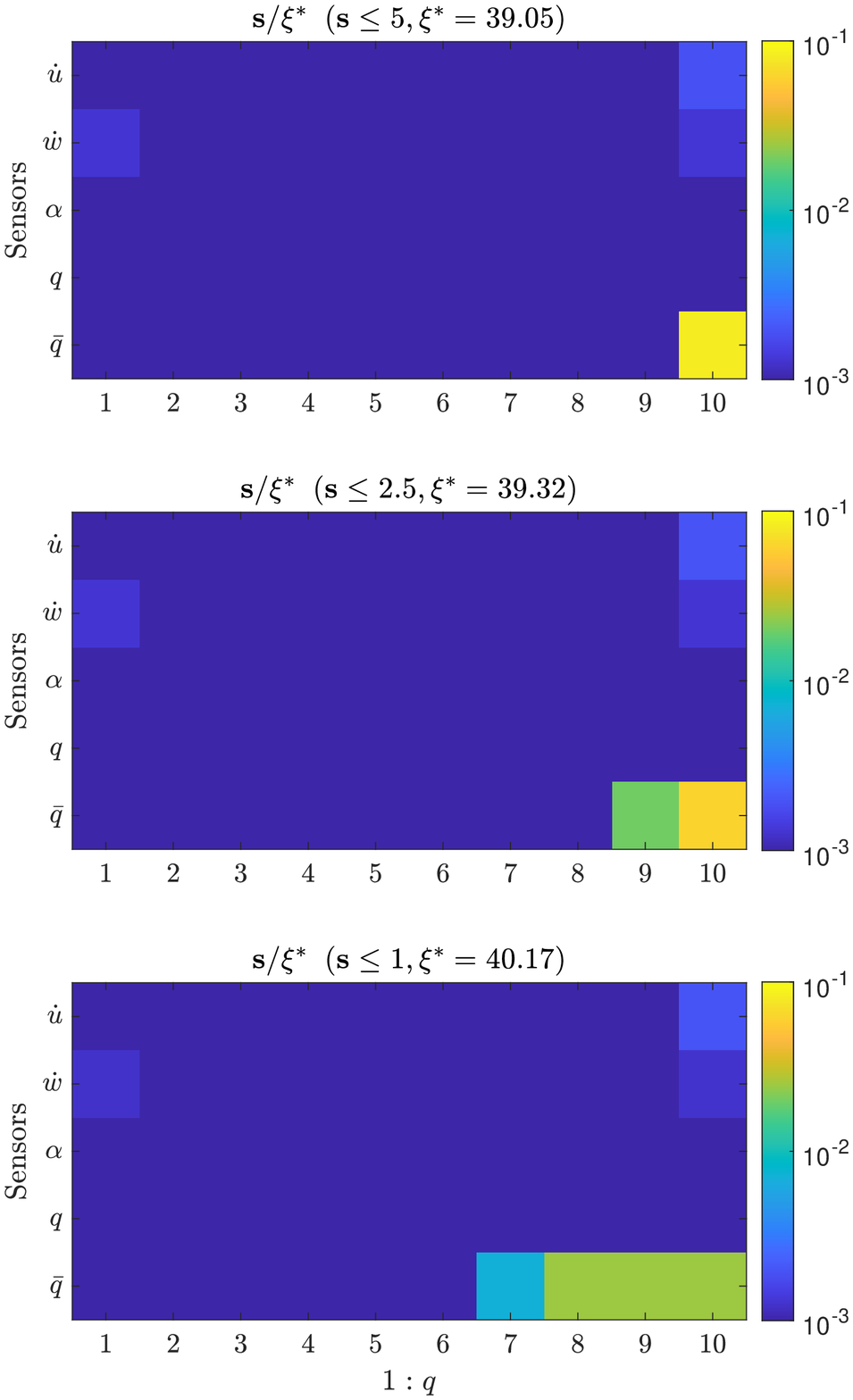}
\caption{Optimal scaled precisions for high sensing rate, guaranteeing $\trace{\M_x\P_\infty\M_x^T}\le\gamma_d$ for $\gamma_d:=0.1$.}
\figlabel{mr_reshape}
\end{figure}

In this example, the $5$ physical sensors in \eqn{sensors} are treated as virtual sensors over $10$ time steps and are assigned an unknown precision. Thus, there are $50$ virtual sensors. Optimization in \eqn{thm2}, results in the sparse precisions shown in \fig{mr_reshape}. The $y$-axis are the five physical sensors, and the $x$-axis are times steps from $1$ to $q$. The heat-map shows the precisions of the $5$ sensors across the $10$ time steps. The three panels in \fig{mr_reshape} are solutions with $\vo{s}_\text{max} = 5, 2.5, 1$ respectively, and scaled by $\xi^\ast$ determined by Algorithm \ref{alg:scaling}. They have the same required accuracy, defined by $\gamma_d = 0.1$. In the top panel, we see that the required precisions for sensors with $1000$ Hz are much lower than the results shown in \fig{sr_moresparse}, which is for $100$ Hz sensing. Therefore, less precise data at a higher rate can achieve the same accuracy. It is also interesting to note that the sensors with nonzero precision are different in the two cases. In the $100$ Hz example, the angular velocity sensor is the most precise, followed by the dynamic pressure sensor. Other sensors have very low precision. In the $1000$ Hz example, the angular velocity sensor and angle of attack sensors have very low precisions. Still, acceleration measurements have relatively higher precisions, with the dynamic pressure sensor the most precise. We also observe that the dynamic pressure sensor plays an important role in both cases, and the angle-of-attack sensor has very low precision in both cases.

It is also interesting to note that the sensor values at the beginning and the end of the 10 time-step window have higher precision. However, as $\vo{s}_\text{max}$ is reduced, we observe that intermediate values of $\bar{q}$ are needed to achieve the same accuracy. We can infer from this observation that if available sensor precision is low, we can achieve a higher estimation accuracy by fusing data at a higher rate. Theorem \ref{thm:2}, determines data from which sensors are needed at a higher rate and the corresponding precisions to achieve this accuracy. This is very useful from a real-time scheduling perspective because from \fig{mr_reshape} we can determine exactly when to poll the sensors. This optimizes sensor polling, reduces the associated delays, and improves real-time schedulability. 

\subsection{Time-Varying System: Satellite Tracking Problem}
This example applies theorem \ref{thm:1} to a linear \textit{time-varying} discrete-time system. Here we consider the problem of determining the optimal sensor precision for tracking a space object with the required accuracy. 

\subsubsection{Model}
Here we consider a simple satellite dynamics model \cite{junkins2009analytical} with $J_2$ perturbation given by
\begin{subequations}
\begin{align}
\ddot{r} &= -\frac{\mu_E}{r^2} + \dot{\theta}^2r + \frac{3J_2}{2r^4}\left(3\sin(\theta)^2-1\right),\\
\ddot{\theta} &= -\frac{2\dot{\theta}\dot{r}}{r} - \frac{3J_2}{r^4}\cos(\theta)\sin(\theta).
\end{align}
\eqnlabel{satDyn}
\end{subequations}
where $r$ is the distance of the satellite from the centre of the orbit and $\theta$ is the angular position of the satellite in the orbit.
Length and time in the dynamics are normalized using $R_E$ (radius of Earth), and $T_p$ (time for one orbit) respectively. The nominal trajectory is the solution of \eqn{satDyn} with normalized initial condition
\begin{subequations}
\begin{align}
 r_0 &= \frac{R_E+h}{R_E} = 1.0533, \\
\dot{r}_0 &= 0, \\
\theta_0 &= 0,\\
\dot{\theta}_0 &= \left(\frac{V_\theta T_p}{R_E}\right)\frac{1}{r_0} = 6.2832. 
\end{align}
\end{subequations}
The parameters necessary to simulate the system are provided in table \ref{table:satData}.
\begin{table}[h!]\begin{center}\hrule
\begin{tabular}{ll}
$R_E = 6378.1363$ km & $\mu_E = 398600.4415$ km$^3$/s$^2$\\
$T_p = 5.48 \times 10^3$ s & $J_2 = 1.7555\times10^{10}$ km$^5$/s$^2$\\
$V_\theta = 7.7027$ km/s & $h$ = 340 km
\end{tabular}\hrule\vspace{1mm}
\caption{Parameters in the satellite dynamics model.}
\label{table:satData}
\end{center}
\end{table}
Equation \eqn{satDyn} is linearized about the nominal trajectory to obtain a continuous-time periodic system. We augment the linear model with process noise,
to account for the effects of sporadic thrusts that are necessary for orbital station keeping. The augmented model is given by,
\begin{align}
\dot{\x} = \A(t)\x + \B\w(t), \eqnlabel{pertDyn}
\end{align}
where $\x := \begin{bmatrix}r & \dot{r} & \theta & \dot{\theta}\end{bmatrix}^T$ is the state vector, $\vo{w}(t) := \begin{bmatrix}w_r(t) & w_\theta(t)\end{bmatrix}^T$ is a zero-mean Gaussian random process,
\begin{subequations}
\begin{align}
\A(t) := \begin{bmatrix} 0 & 1.0 & 0 & 0\\ a_{21}(t) & 0 & a_{23}(t) & 12.59\\ 0 & 0 & 0 & 1.0\\ a_{41}(t) & -12.21 & a_{43}(t) & 0 \end{bmatrix}, & \B := \begin{bmatrix} 0 & 0 \\ 1 & 0 \\ 0 & 0 \\ 0 & 1 \end{bmatrix},
\end{align}
\end{subequations}
with
\begin{align*}
a_{21}(t) &= 0.416\cos(12.4t) + 126.4,\\
a_{23}(t) &= 0.2113\sin(12.4t),\\
a_{41}(t) &= 0.2774\sin(12.4t),\\
a_{43}(t) &= -0.1408\cos(12.4t).
\end{align*}
In this example, we assume the mass of the satellite is $100$ kg, and the satellite sporadically applies maximum of 1mN of thrust for orbital station keeping. The normalized accelerations due to these thrusts are modeled as zero-mean Gaussian random processes $w_r(t)$ and $w_\theta(t)$, with $\Exp{w_r w_r^T} = \Exp{w_\theta w^T_\theta} = 0.0471^2$. \Fig{ltvp_olpStat} shows the propagation of mean $\mub(t)$ and variance $\Sigb(t)$ for the time-varying linear system, with 
\begin{subequations}
\begin{align}
\mub(t_0) &:= \begin{bmatrix}50/R_E \\ 0 \\ 0 \\ 0 \end{bmatrix}, \text{ and } \\
\Sigb(t_0) &:= 0.01\times\diag{\mub(t_0)} \eqnlabel{stat:icCov}.
\end{align} \eqnlabel{stat:ic}
\end{subequations}
The evolution equation for $\mub(t)$ and $\Sigb(t)$ are given by
\begin{align}
\dot{\mub}(t) &= \A(t)\mub(t),\\
\dot{\Sigb}(t) &= \A(t)\Sigb(t) + \Sigb(t)\A^T(t) + \B\Q\B^T, \eqnlabel{covProp}
\end{align}
where $\Q := 0.0471^2\times\I_{2}$.

\begin{figure}[h!]
\includegraphics[width=0.45\textwidth]{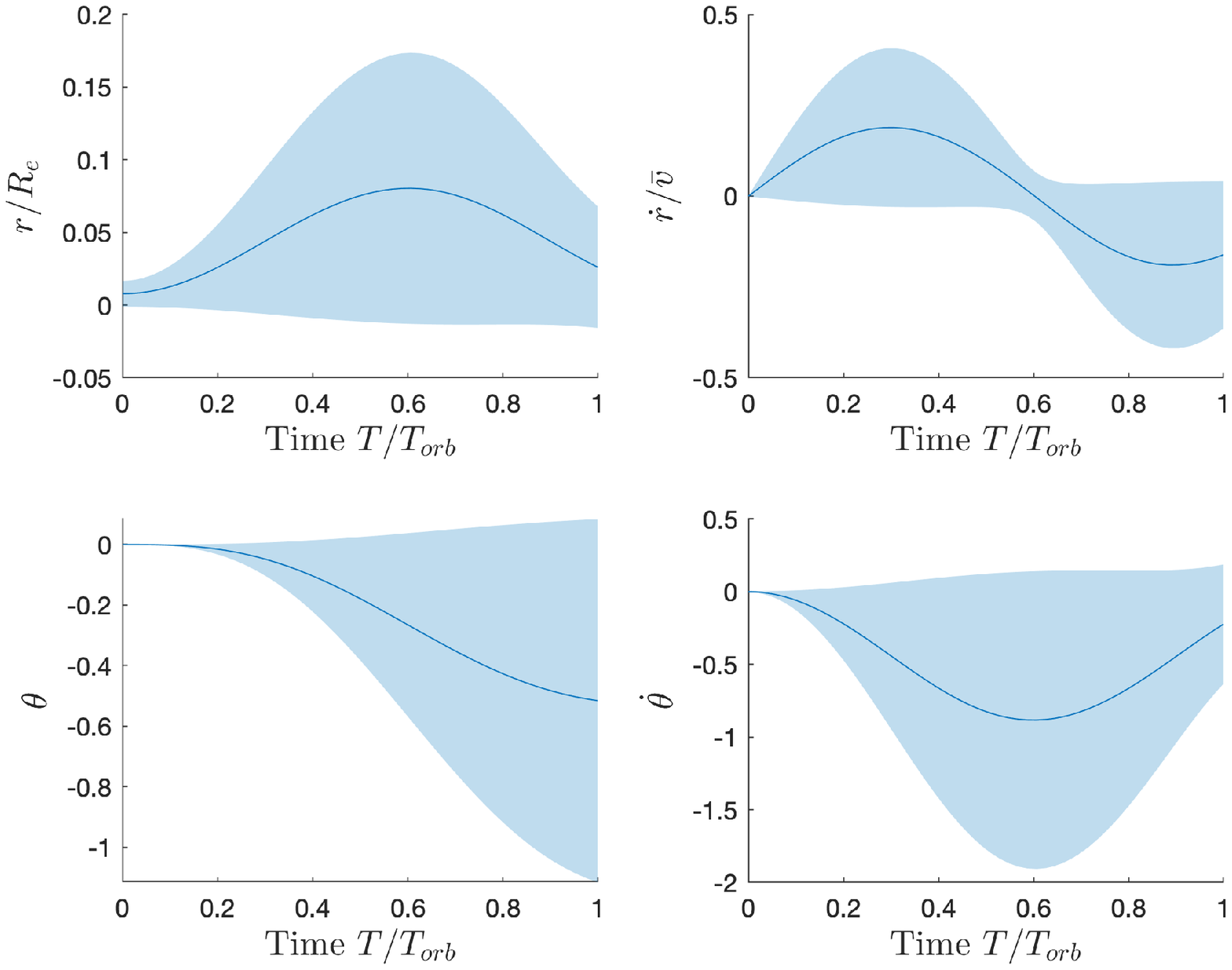}
\caption{Uncertainty propagation with the linear time-varying periodic system. Solid line shows the evolution of the mean perturbation, and the shaded region shows $\mub_i \pm \sqrt{\Sigb_{ii}}$ for $i=1, 2, 3$ and $4$.}
\figlabel{ltvp_olpStat}
\end{figure}
We discretize the normalized time interval $[0,1]$ with $dt = 0.1$, resulting in the temporal grid $\{t_k\}$, where $t_k:=kdt$. We assume that measurements are available at these times. The dynamics in \eqn{pertDyn} is discretized over $\{t_k\}$, and is given by
\begin{align}
\x_{k+1} = \A_d(t_k)\x_k + \w_k,
\end{align}
where
\begin{align}
\A_d(t_k) &:= \vo{\Phi}(t_{k+1},t_{k}),\\
\w_k &:= \int_{t_k}^{t_{k+1}} \vo{\Phi}(\tau,t_{k})\B\w(\tau)d\tau, \eqnlabel{w_k}
\end{align} 
and $\vo{\Phi}(\cdot,\cdot)$ is the state-transition matrix, which is obtained by numerical integration of the fundamental matrix. 

It is easy to verify that if $\Exp{\w(t)} = 0$, then $\Exp{\w_k} = 0$. Therefore, $\w_k$ is a zero-mean random process. To determine the optimal precision, we need to quantify $\Q_k := \Exp{\w_k\w^T_k}$, which is difficult to determine from \eqn{w_k}. Instead, we use the covariance $\Sigb(t)$, determined by solving \eqn{covProp}, and the discrete-time covariance propagation equation, to determine the time-varying $\Q_k$. It is given by
\begin{align}
\Q_k := \Sigb(t_{k+1}) - \A_d(t_k)\Sigb(t_k)\A^T_d(t_k).
\end{align} 
 
\subsubsection{\textbf{Example 3:} Optimal Sensor Scheduling}
In this example, we consider a set of 10 laser-ranging sensors located on the surface of the Earth, at angular positions $\theta(t_k)$. For the periodic system described above, the objective is to determine the optimal sensor precisions such that $\trace{\Sigb^+(t_k = 1)}\le \gamma_d$, given $\Sigb^-(t_k = 0)$. Here we apply theorem \ref{thm:1} to determine the optimal sensor precisions, which are shown in \fig{satPrecisionsPolar} for various values of $\vo{s}_\text{max}$. 

The optimization is done with $\Sigb(t_k = 0) = \Sigb(t_0)$, and $\gamma_d = 0.1\times\trace{\Sigb^-(t_k = 1)}$, where $\Sigb^-(t_k = 1)$ is the prior obtained at $t_k=1$. It is obtained by propagating $\Sigb(t_k = 0)$ using \eqn{covProp}. Variance $\Sigb(t_0)$ is defined in \eqn{stat:icCov}. The value of $\gamma_d$ specifies that we want the trace of the posterior to be 10\% of the prior at $t_k=1$.

\begin{figure}[h!]
\includegraphics[width=0.45\textwidth]{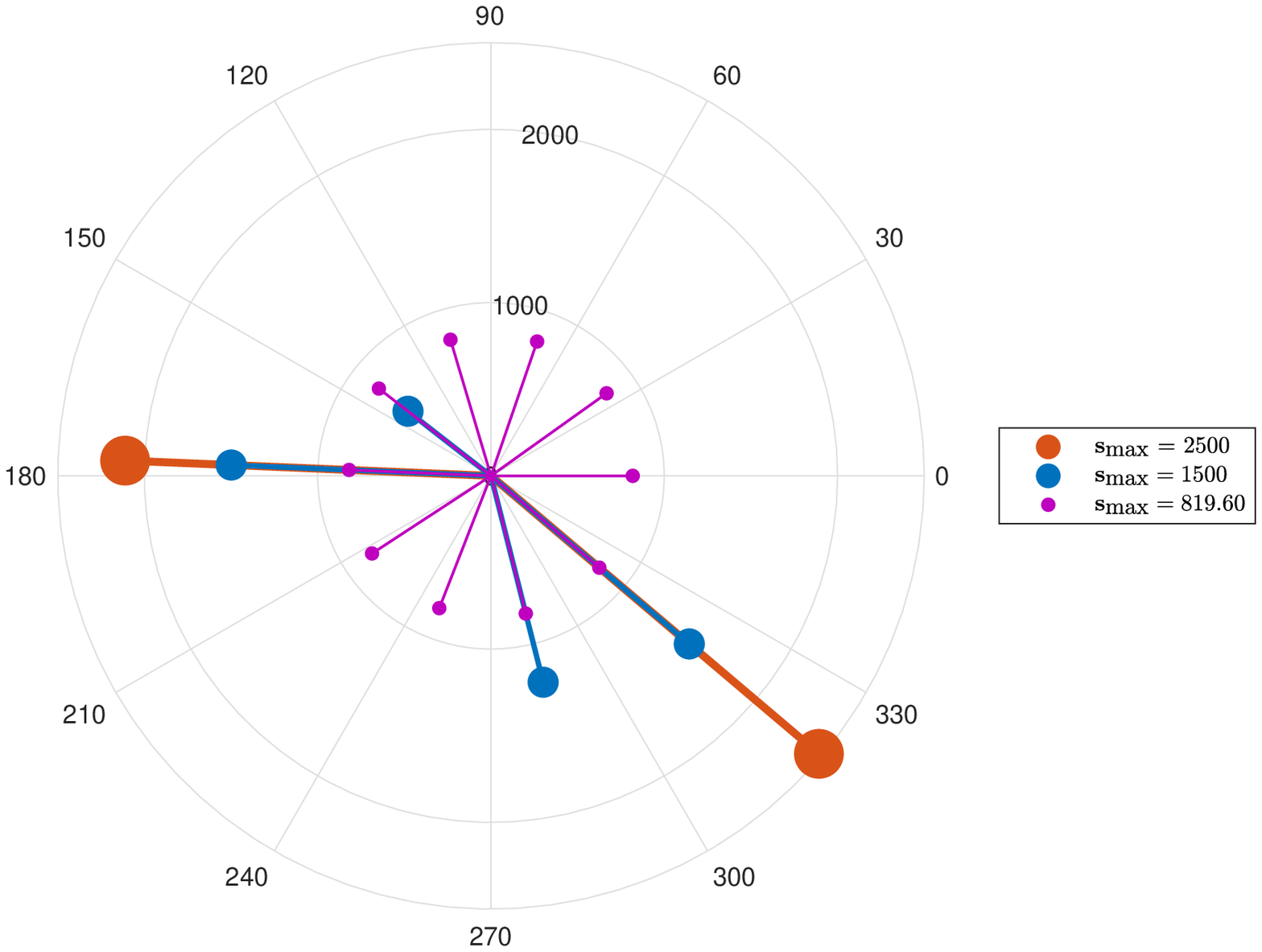}
\caption{Optimal precisions for $\vo{s}_\text{max} = 2500, 1500, 819.60.$}
\figlabel{satPrecisionsPolar}
\end{figure}

From \fig{satPrecisionsPolar}, we see that there is a tradeoff between sensor precision and sensing frequency. With low precision sensors ($\vo{s}_\text{max} = 819.60$), we need to sense the satellite at all the sites to guarantee $\trace{\Sigb^+(t_k = 1)}<\gamma_d$. As $\vo{s}_\text{max}$ is increased, the sensing becomes more sparse. With $\vo{s}_\text{max} = 2500$, we only need to get two range measurements, in order to estimate the state at $t_k=1$ with the required accuracy.

This is significant for large-scale spatio-temporal sensing, and especially for tracking space objects. Currently, there are about $500K$ space objects, and only $30K$ are tracked. The number of sensing sites is significantly lower than that. Therefore, by applying theorem \ref{thm:1}, it is possible to determine the sparse sensing schedule for each object, thereby increasing the ability to track more objects with guaranteed accuracy. 

Precision in laser-ranging sensors is determined by the energy of the laser beam and the reflectivity of the object being sensed. A lower value of precision implies lower energy requirements for sensing the object, which results in optimal sensor designs. Modern satellites have reflectors, which allows precise sensing with low powered lasers. Some space objects (e.g., asteroids, etc.) have poor reflectivity, which fundamentally limits how accurately it can be sensed. This limit can be incorporated using the variable $\vo{s}_\text{max}$.

\section{Summary \& Conclusion}

This paper presented convex optimization problem-formulations to determine optimal sensor precisions that guarantee a specified estimation accuracy. The formulation is presented in a general multi-rate sensing framework, with linear time-varying discrete-time system dynamics. Optimality is achieved by minimizing sensor precisions, subject to the upper bound on the estimation error, as defined in the discrete-time Kalman filtering framework. Since the minimization of precisions is done with respect to the $l_1$ norm, the proposed optimization framework can also be used to determine sparse sensing architectures. This will be valuable in the design of large-scale sensor networks. We have shown the proposed theory's engineering value by applying it to realistic flight mechanics and astrodynamics problems. 

\section{Acknowledgements}
This research has been supported by the National Science Foundation grant \#1762825.


\bibliographystyle{unsrt}
\bibliography{double}

\end{document}